# Electron irradiation: from test to material tailoring

A. Alessi, O. Cavani, R. Grasset, H.-J. Drouhin, V. I. Safarov, and M. Konczykowski

LSI, CEA/DRF/IRAMIS, CNRS, Ecole polytechnique, Institut Polytechnique de Paris, 91120 Palaiseau, France.

In this article, we report some examples of how high-energy electron irradiation can be used as a tool for shaping material properties turning the generation of point-defects into an advantage beyond the presumed degradation of the properties. Such an approach is radically different from what often occurs when irradiation is used as a test for radiation hard materials or devices degradation in harsh environments. We illustrate the potential of this emerging technique by results obtained on two families of materials, namely semiconductors and superconductors.

**Introduction.** Decades of research have shown that irradiation is capable of strongly modifying the physicochemical properties of a material. Such effects can lead to degradation of materials and associated devices and set limits on their use for specific applications [1-7]. Along their path within matter, electrons can interact in many ways [8]. A large part of their energy is transferred to the material by interaction with electrons: these collisions are responsible for ionization phenomena. At the same time, electrons can also undergo collisions with the nuclei, inducing their displacement from regular crystal sites. The consequence of this process is the generation of a vacancy and of an interstitial atom. Such process is defined as non-ionizing energy loss and it determines the displacement damage [6, 9].

Displacement damage contributes to the degradation of material properties and device performance in harsh environments. This is the case for electronic devices [6] and especially for solar cells used for space applications [5, 10]. In this domain, electron irradiation is a widely used tool to test the radiation response of the solar cells and guarantee energy production at a sufficient level throughout the satellite's mission. Over time, electronic irradiation has been directed towards the search for materials more resilient to radiation and for the production of solar cells resistant to the extreme conditions encountered in space [11]. Thus, irradiation has been more and more involved in research projects, a trend that continues and grows. Similar examples of this type of use of electron irradiation can also be found in other research fields [1]. Irradiation by high-energy electrons (HEE) is truly different from other irradiation techniques: indeed, due to the small electron mass, energy transfer to heavier nuclei remains quite tiny. Protons or other heavy particles can induce similar damage processes, but these particles transfer a very important amount of energy thus the first collision can generate a cascade of secondary events, leading to the production of complex and extended defects [1, 6, 9, 12]. On the contrary HEE irradiation generates mostly isolated point defects, namely Frenkel pairs consisting of vacancy and interstitial atom [13]. Then, when the energy barrier for the migration of the interstitials is lower than that of the vacancies, what happens frequently, the interstitials migrate very easily at room temperature and are trapped at dislocations and/or surface leaving behind only vacancies. Their concentration is controlled exclusively by fluence, i.e., the number of incident electrons per unit area which is conveniently measured by the charge per $cm^2$ ($C.cm^{-2}$).

It is important to note that the resulting disorder is truly random, which is not so easy to achieve in other ways. Electrons have a large electron penetration depth (reaching several mm at 2 MeV) so that bulk samples can be irradiated homogeneously. The induced point defects create localized energy levels in the electronic band structure and also act as scattering centers. Localized levels crucially alter semiconductor properties whereas scattering centers are of primary importance for superconductors. By adjusting fluence, they can be introduced progressively and in a controlled way, allowing fine tuning of material properties.

As a consequence, nowadays, there are activity areas in which HEE irradiation has become a tool for shaping materials with specific properties and a real step in technological improvement.

Hereafter, we present two examples of this emerging trend of HEE irradiation. One is the control of the Fermi level position of wide bandgap materials while the other is the controlled introduction of disorder into superconducting materials to explore phase transitions.

The first example will be focused on $Ga_2O_3$ that is an emerging ultra-wide bandgap semiconductor with potential applications in high power electronics [14]. $Ga_2O_3$ is supposed to be a radiation hard or a radiation tolerant material so it also a good candidate for extreme environments like space [14]. $Ga_2O_3$ present, however, some limitations as high thermal resistance and the difficulty of producing p-type materials using dopant [14].

The second example will be focused on an unconventional superconductor ($Ba(FeAs_{1-x}P_x)_2$) in which it is possible to use the presence of impurities, so also of defects, to increase the disorder and to probe its pairing symmetry. Radiation effects on this material have been sufficiently studied to allow strong conclusions [15].

**Ga$_2$O$_3$: an example of semiconductor doping with point defects.** Generation of point defects may give rise to energy levels in the bandgap of a semiconductor [16, 17]. Specifically, vacancies can act as donors or acceptors, thereby affecting the electrical conductivity of the materials. In this way, high-energy-electron irradiation can be considered as a specific doping process, the dopant concentration being easily controlled by fluence. For example, in germanium these defects act as acceptor-like centers and their generation increases the resistivity of the *n*-type material, possibly leading to conversion to a *p*-type material [1, 18]. Similarly, in CdGeAs$_2$, conversion from *n*- to *p*-type conductivity was reported after HEE irradiation [19]. In compounds the situation is subtle since the vacancies in different sub-lattices have different electronic properties. It is therefore important to know the relative concentration of vacancies produced in the different sub-lattices of the crystal. During an electron-nucleus collision, the electron can transfer to a much heavier nucleus only a tiny amount of its energy [20-22]. In particular, the maximum transfer of energy $E_{Imax}$ (for relativistic electrons) can be calculated by equation (1):

$$E_{Imax} = \frac{2(E_e + 2m_0 c^2)}{M_I c^2} E_e, \qquad (1)$$

where $E_e$ is the energy of the incoming electron, $m_o$ is the electron rest mass ($m_0 c^2 = 0.511$ MeV), and $M_I$ is the mass of the target nucleus. By way of illustration, Fig. 1 shows a contour plot of the transferred energy to Fe atoms (in Ba(FeAs$_{1-x}$P$_x$)$_2$) and makes clear that HEE irradiation is a reliable way to generate point defects in materials (after Ref. 15). Using Eq. (1), for $^{69}$Ga, electrons with 0.5 MeV, one obtains a maximum energy transfer of about 30 eV. Clearly, a nucleus can be displaced from its regular site only if the transferred energy exceeds a critical value related to the energy of the chemical bonds with neighboring atoms $E_b$. In the following of this section, we will use recent data obtained for Ga$_2$O$_3$ as an example of how electron irradiation can be used as a doping technique [23, 24].

From conductivity measurements on irradiated samples, it has been deduced in Ref. 24, that the value of the gallium binding energy can be taken as 25 eV or so. From Eq. (1), it follows that only electrons with energies higher than the threshold energy $E_t \sim 0.5$ MeV are able to transfer to a gallium nucleus the required amount of energy and displace it from its position on the lattice. For a lighter oxygen atom, the maximum energy transfer is larger, which implies (for similar or lower $E_b$) a lower energy threshold for the displacement.

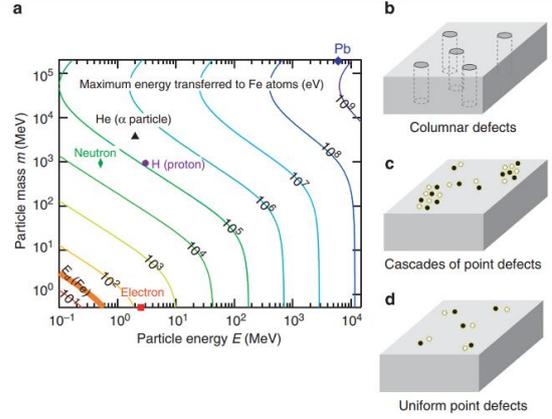

Fig. 1: Particle irradiation and created defects. (a) Contour plot of maximum energy transferred to Fe atoms (recoil energy) for impinging particles with rest mass *m* and incident energy *E* (see Eq. 1). Typical threshold energy $E_d$ for the displacement of Fe atoms from the lattice sites is marked by thick orange line. Commonly used energies for electrons (red square), neutrons (green diamond), protons (purple circle), α particles (black triangle) and heavy-ions (Pb, blue diamond) irradiation are indicated. (b–d) Schematic illustrations for different disorder morphologies created by particle irradiation. Columnar defects can be created by heavy-ion irradiation (b). Particle irradiation involving relatively large energy transfers tends to produce cascades of point defects (c). Electron irradiation is the most reliable way to obtain uniform point defects (d). Figure first published in Ref. 15.

Then, to evaluate the efficiency of vacancy generation, we must consider the cross section for Rutherford scattering which is proportional to the square of the nucleus charge $(Ze)^2$ [22]. This latter expression implies that highly-charged heavy nuclei are more exposed to collisions with incoming electrons than lighter ones.

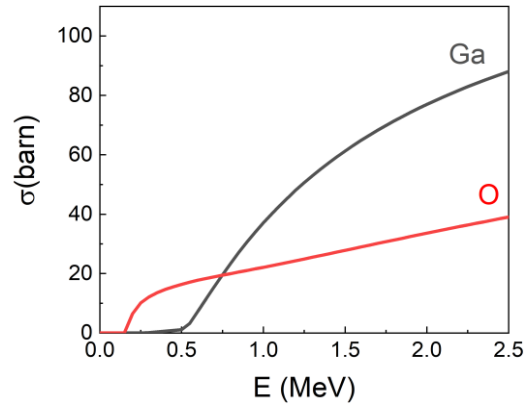

Fig. 2: SECTE simulation of O or Ga vacancy cross-sections as a function of the energy of the incoming electrons. For the sake of simplicity, the typical value $E_b = 25$ eV has been taken for Ga and O atoms, whatever the relevant sub-lattice.

The calculation of the total effective cross-section $\sigma$ for atom displacement, considering both above-mentioned processes, requires cumbersome numerical computations. For this purpose, a numerical code, the SECTE software has been developed at LSI [25]. Figure 2 presents the variation of $\sigma$ with the energy of the incoming electrons.

At low electron energies, electron irradiation essentially affects the oxygen atoms due to their lower thresholds for displacement. In contrast, regarding higher-energy electrons, $E_e >> E_{tGa}$, collisions with heavier gallium atoms dominate compared to that with oxygen atoms. As follows from Fig. 2 with a proper choice of the electron energies, we can preferentially create vacancies in one of the crystal sub-lattices constituted by given chemical species, while fluence tunes the concentration. As mentioned above, the vacancies in different sub-lattices have different electronic properties. In β-$Ga_2O_3$ the oxygen vacancies act as donor-like centers [26] and their presence is often considered to cause $n$-type conductivity in unintentionally doped crystals. On the contrary, the Ga vacancies are considered to play the role of acceptors [27]. Based on the data in Figure 2, irradiation with electrons of 2.5 MeV should create mostly acceptor-like gallium vacancies. To prove this, we performed irradiation at 20 K on gallium oxide initially doped with Sn up to the concentration $N_d = 2 \times 10^{18}$ cm$^{-3}$ with the SIRIUS electron accelerator [23, 24]. Figure 3 shows the variation of the concentration of conduction electrons with fluence measured at room temperature after irradiation at 20 K. The observed linear reduction of the conduction electrons concentration confirms the ability of electron irradiation to tune the electric properties of β-$Ga_2O_3$ by the introduction of acceptors whose concentration increases linearly with the irradiation dose. The slope of the experimental curve indicates that the concentration of acceptors varies as $N_a = 1.24 \times 10^{16}$ (1/(cm×mC)) × F (mC/cm$^2$).

The concentration of gallium vacancies can be evaluated by the equation $N_{Ga}= \sigma \times$[Ga]$\times$F (where σ is cross-sections, [Ga] gallium atom concentration and F is the fluence expressed as number of electron for cm$^2$). By comparing the values obtained by this second equation with $N_a$ we found comparable values of acceptors and gallium vacancies. So, in first approximation for the data reported here it is possible to explain the radiation effect by $N_{Ga}$ generation. Differences between $N_a$ and $N_{Ga}$ could be due to the fact that: i) the estimated σ are useful to understand which is the main type of vacancy generated as a function of the beam energy but they can be a roughly estimation of the real cross section; ii) oxygen vacancies, having opposite effect, are generated too;

iii) partial annealing of vacancy going from 20 K to 300 K (thermal stability under investigation).

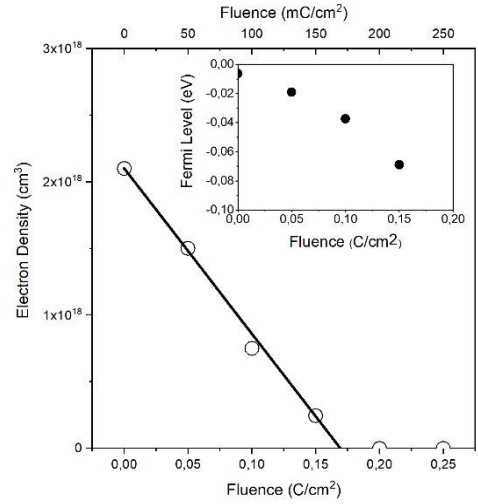

Fig. 3: Carrier concentration (conduction electrons) versus fluence evaluated from Hall measurements in β-$Ga_2O_3$ after irradiation with 2.5 MeV electrons. In the inset, we report the calculated Fermi level position with respect to the conduction band. The concentration of conduction electrons arises from the imbalance between donor and acceptor concentrations $n = N_d-N_a$.

The variation of the carrier concentration can be related to the Fermi level position. The inset in Fig. 3 shows the shift of the Fermi level with irradiation. This downward shift causes the modification of other properties of the crystal. In fact, after electron irradiation of unintentionally doped $n$-type gallium oxide samples new lines appear in optical and EPR spectra [24]. These lines correspond to $Cr^{3+}$ and $Fe^{3+}$ ions. Chromium and iron are known to be present as impurities in $Ga_2O_3$ [28], however, in $n$-type samples with a high position of the Fermi level, the chromium and iron ions accept one more electron and turn to less charged ions like $Cr^{2+}$ and $Fe^{2+}$ [24, 29]. While $Cr^{3+}$ and $Fe^{3+}$ replacing $Ga^{3+}$ ions are "neutral" with respect to the host lattice, the doubly charged impurities may be considered as negatively charged donor-like centers where an extra electron is bounded to the Cr or Fe impurities to achieve a more stable electronic configuration of the d shell (see e.g. Klechkowski's and Hund's rules). After proper irradiation, the consequent downward shift of the Fermi level returns them to triple-charged "neutral" states that are detectable in optical and EPR spectra. Other important modifications of the EPR spectrum are the disappearance of the shallow donor resonance and the appearance of a more complex signal [23,24] that was attributed to Ga vacancy (in a tetrahedral site and in a double negative charge state) [30] or to some complex like $V_{Ga}$–$Ga_i$–$V_{Ga}$ [30, 31].

As previously mentioned, $Cr^{3+}$ is detected in EPR which implies also the modification of the emission properties. Figure 4 illustrates an example of $Cr^{3+}$ photoluminesce (PL) spectrum recorded on an unintentionally doped ($1.5 \times 10^{17}$ cm$^{-3}$) $Ga_2O_3$ sample after irradiation at 80 mC/cm². A more detailed investigation of the emission modifications can be found in ref [24].

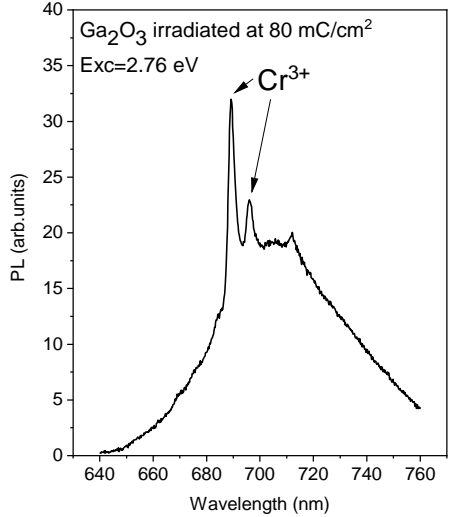

Fig. 4: PL spectrum recorded on a $Ga_2O_3$ sample after irradiation.

Irradiation with electrons can be also used to modify the transmission properties. In fact, in some materials, e.g., $CdGeAs_2$ and $ZnGeP_2$ which are used for their near-infrared nonlinear optical properties [19, 32-35], defect conversion and Fermi level shift lead to improved transparency, proving that HEE irradiation can adjust material properties to meet application requirements.

**Probing pairing symmetry in superconductors by disorder effect.** The discovery of novel families of superconducting materials triggered interest in testing the mechanism of formation of Cooper pairs. While in high-$T_c$ cuprates phase-sensitive tunneling measurements gave direct evidence of d-wave pairing symmetry, such experiments proved impossible on strongly correlated materials and in particular on iron based superconductors. After that, scattering effect by non-magnetic impurities has become the touchstone for the study of the nonconventional pairing mechanism in those materials. Historically, Anderson predicted a total insensitivity of critical temperature and superconducting gap to non-magnetic scattering centers in single band isotropic s-wave superconductors [36]. Later, Abrikosov and Gor'kov refined the model of superconductor including scattering and predicting the depression of $T_c$ as a function of the scattering rate [37]. The standard experimental protocol, used to distinguish the conventional pairing mechanism from the exotic pairing mechanism, makes use of the relation between critical temperature and normal-state resistivity at various degrees of irradiation-induced disorder. The rate of depression of $T_c$ with disorder, expressed quantitatively by the increment of resistivity in the normal state, is compared with that expected from the Abrikosov - Gor'kov formula:

$$\ln(t_c) = \psi\left(\frac{1}{2}\right) - \psi\left(\frac{1}{2} + \frac{2g_p}{t_c}\right), \quad (2)$$

where $\psi$ is the digamma function, $t_c = \frac{T_c}{T_{co}}$ is the reduced critical temperature, and $g_p$ the pair-breaking parameter. The latter is directly related to the dimensionless scattering rate $g^\lambda$

$$g^\lambda = \frac{\hbar \Delta\rho}{2\pi k_B \mu_0 T_{c0} \lambda_0^2}, \quad (3)$$

where $\Delta\rho$ stands for the increment of normal state resistivity induced by irradiation, $\lambda_o$ is the zero temperature London penetration depth, and $T_{c0}$ the initial critical temperature [38]. Comparison of the observed depression rate of $T_c$ plotted against the normal state resistivity - proportional to the irradiation dose - with that calculated with Eq. (2) is thus used for identification of the paring symmetry [21]. Figure 5 gives an example of this kind of investigation in $Ba(FeAs_{1-x}P_x)_2$.

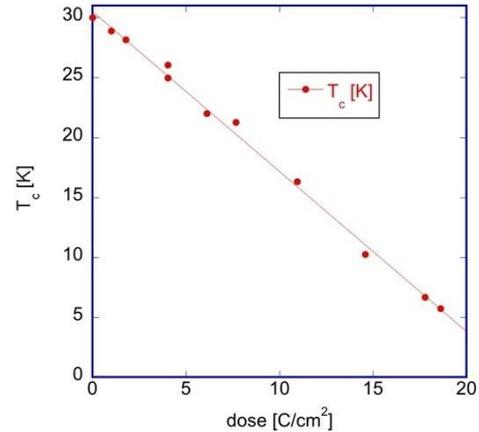

Fig. 5: Evolution of the critical temperature with the irradiation dose observed in optimally doped 122-family iron-based superconductor $Ba(FeAs_{1-x}P_x)_2$. A linear depression with no sign of saturation extends down to 1/6 of the initial value of $T_c$. Data obtained from resistivity measurements after irradiation at SIRIUS.

The data show an impressive linear behavior over a wide temperature range, that is different from what has been observed after proton irradiation in $Ba(Fe_{1-x}Rh_x)_2As_2$, where a slower $T_c$ depression was measured with saturation at $T_c/T_{co} \sim 0.85$ [39]. This discrepancy apparently stems from the fact that irradiation with HEE at cryogenic temperature and irradiation with protons do not produce the same type of disorder: HEE generates a genuine random disorder consisting of point defects while, beyond a given dose, proton irradiation yields complex extended defects (see Fig. 1). This compromises the interpretation in terms of superconducting phase transition given in Ref. 39, based on theoretical prediction of Ref. 40.

More in-depth information can be obtained from the temperature dependence of the London penetration length $\lambda$ which is related to the superfluid density $n_s$ by the simple relation: $\lambda \propto n_s^{-1/2}$. Thus, the temperature variation of $\lambda$ is the fingerprint of the topological structure of the superconducting gap, which varies with disorder. Accordingly, the evolution of $\lambda$ versus $T$ in samples irradiated at different fluences reveals disorder driven evolution of functional form of temperature dependence of London penetration depth in optimally doped iron based superconductor Ba(FeAs$_{1-x}$P$_x$)$_2$ (see Fig. 6). As reported in Ref. 15, pristine sample manifests linear dependence, indicative for presence of line of nodes in superconducting gap. Upon introduction of moderate disorder, it turns to exponential variation characteristic for fully gaped superconductor. At high irradiation dose and significant reduction of critical temperature, variation of $\lambda$ become proportional to $(T/T_c)^2$ predicted for fully gaped superconductor with disorder induced mid-gap states.

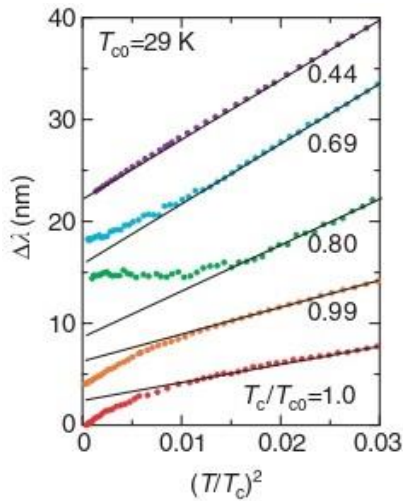

Figure 6: Effect of electron irradiation on the low-temperature penetration depth in BaFe$_2$(As$_{1-x}$P$_x$)$_2$ single crystals. Change in the magnetic penetration depth $D\lambda$ plotted against $(T/T_c)^2$. Each curve is shifted vertically for clarity. Lines are the $T^2$ dependence fits at high temperatures; after Ref. 15.

**In-situ/online experiments**. Present developments have introduced new instrumentation requirements regarding irradiation platforms. This particularly concerns the demand for experimental techniques in situ (without moving the sample, but with the beam turned off) or online (during irradiation), as well as irradiation at different temperatures or different fluxes. One of the main parameters that affects the radiation effects is temperature, because after formation, defects can migrate and then recombine or interact to form new defects.

Such processes are temperature dependent and impose the use of low temperature in-situ or even online experiments to understand the basic mechanisms. Furthermore, depending on the environment, temperature can be different from the room temperature, so that a better understanding of the real response of devices (i.e., under operation conditions) requires the control of the irradiation temperature. Nowadays, at SIRIUS there are two different low-temperature irradiation lines. One is used for in-situ testing of solar cells for space applications down to 100 K [41, 42] and belongs to the category of the advanced test equipment. The other enables to perform irradiation and in-situ resistivity tests at 20 K, being often used to irradiate superconductors and, in general, to limit unsuitable phenomena related to thermally activated processes during irradiation. This latter line, named Cryo1, was born for studies of fundamental physics and proved to be a tool for adjusting the level of doping of materials, thus becoming part of a new perspective for the use of irradiations. In order to improve in-situ characterization, a third irradiation line allowing to perform irradiations down to 5 K coupled with Hall effect and electron paramagnetic resonance (EPR) experiments is under development.

Although we have mainly discussed defects related to displacement damage, in-situ and online experiments also concern ionizing effects and the characterization of materials in which radiolytic effects are relevant. In fact, electron-hole pairs generation is followed by complex series of processes involving transient defects and/or trapping in impurities, pre-existent or radiolytic defects, and diffusion processes which are largely dependent on temperature. In this context, irradiation and investigation at 5 K can be crucial to understand metastable defects, because their unstable nature can be significantly reduced.

Hall effect is a standard tool used to obtain information on the carrier concentration and doping type in a material. In-situ low temperature Hall effect measurements would make it possible to obtain these properties without any undesirable annealing of the defects created.

In situ EPR experiments at around 5 K are an excellent prospect for electron irradiation research. Indeed, this technique can be applied to many types of materials because electron irradiation induces, through electron or hole trapping and bond breakings, generation of paramagnetic defects that can be metastable and therefore not accessible or difficult to explore with ex-situ experiments. In the past, EPR data recorded at 77 K or 110 K on sample irradiated at 77 K without going back at room temperature [43, 44] have given a great contribution to hole self-trapping identification and characterization in silica, allowing a better evaluation of their impact on satellite applications. In the field of optical fibers, on-line radiation-induced attenuation experiments have brought a wealth of information on the presence of transient absorption that have to be considered for applications and that could not be easily investigated after irradiation because of their post irradiation annealing [3, 45, 46].

**High temperature irradiation:** While low temperature irradiations limit or avoid migration of defects and so their interactions, the opposite take place during high temperature irradiation.

An example of positive effect of performing irradiation at high temperature is the generation efficiency of NV centers (nitrogen-vacancy) [47, 48]. In fact, it was observed that in bulk crystals the coupling of annealing and irradiation produces more defects than a room temperature irradiation followed by annealing [48]. More detailed discussion on the generation of these defects in nanosystems and bulk can be found in ref [47,48]. A crucial point is that during high temperature irradiation the vacancies needed to form the NV are mobile after their creation while in the other case they become mobile only when the thermal treatment is applied. This happens after irradiation when vacancies and other defects are present in high concentrations, which implies a larger possibility that the vacancy interacts with other defects (vacancy clustering as example) instead to interact with a substitutional N atom to form the NV center. Without entering in the discussion of the model, the results are an empirical prove of the possible use of the high temperature irradiation to introduce defect, as NV in the negatively charged state, with a number of peculiar features that make it and the material containing it promise candidates in several application fields [47, 49].

Other possible positive effects can be imagined such as cluster formation or interaction between metastable defects to form other structures that modify the material properties giving it peculiar useful features.

Finally, in ref [50] it can be found an example of the study of graphite-diamond transition as a function of the flux and the temperature of irradiation. Even if the data are related to TEM irradiation that usually have lower beam energy and higher fluxes in this manuscript the energy of the beam is of about 1.25 MeV and low fluxes are considered too. Even in this case displacement damage generated defects play a key role.

**Conclusion.** Irradiation with electrons has a long history and its use and purposes continue to evolve. Today, it aims to become a means to manipulate the properties of matter so that irradiation could be considered as part of the production stage of materials with specific properties. This occurs since the generation of point defects can effectively be used as a tool for shaping material properties rather than being suffered as a detrimental consequence of the interaction between radiation and matter.

In parallel, investigation techniques have evolved to probe the response to radiation under in operando conditions, i.e., increasingly close to those of actual application. Future prospective goes in the direction of low-temperature irradiations coupled with in-situ or even better online experiments.

*\*\*\**

The authors thank the Agence de l'Innovation de Défense, Ministère des Armées, for support under grants 2018 60 0074 00 470 75 01/DIGAO and DIGAO2 and the Agence Nationale de la Recherche (ANR, French national research agency, GOSPELS Projet N°ANR-22-CE50-0017-01) for financial support.